%% file: main.tex
\documentclass[%
 reprint,
 amsmath,amssymb,
 aps,
pra,
]{revtex4-2}

\usepackage{graphicx}
\usepackage{dcolumn}
\usepackage{bm}

\usepackage{placeins} 
 
\usepackage{mathtools}

\input{customCommands}

\usepackage{lipsum}

\begin{document}

\preprint{APS/123-QED}

\title{Translational diffusion coefficients of membrane protein aggregates in free and supported lipid membranes}

\author{Yannick A. D. Omar}
\affiliation{Massachusetts Institute of Technology, Department of Chemical Engineering}
 \email{yadomar@mit.edu}

\date{\today}

\begin{abstract}

There is increasing evidence that numerous membrane proteins can assemble into aggregates that modulate their function and affect many cellular processes such as signal transduction and endocytosis. Here, we present a theoretical description of the instantaneous translational diffusion coefficients of transmembrane protein aggregates on free and supported lipid membranes using Kirkwood-Riseman theory. We find that hydrodynamic interactions within protein aggregates must be accounted for, as neglecting them yields several times lower diffusion coefficients. By deriving hydrodynamic radii for free and supported lipid membranes, we identify effective length scales that accurately characterize aggregate diffusivities in the presence of hydrodynamic interactions. These findings motivate the approximation of an aggregate by its outline and a random particle distribution inside it. We show that this approach provides a practical method to accurately determine aggregate diffusion coefficients when the particle locations cannot be resolved. The results presented in this article have immediate implications for the formation and function of membrane protein aggregates. 

\end{abstract}

\maketitle

\input{content}

\noindent\textbf{Acknowledgments}
I would like to thank Arup K. Chakraborty for helpful feedback and financial support through the grant stated below. I am also grateful to Mehran Kardar for insightful discussions and to Patrick Fleming for pointing me towards his work on convex hull approximations in three dimensions. Research reported in this article was supported by National Institute of Allergy and Infectious Diseases of the National Institutes of Health under award number 2P01AI091580-11A1.

\bibliography{bibliography}
\end{document}

%% file: customCommands.tex
\let\originalleft\left
\let\originalright\right
\newcommand{\leftR}{\mathopen{}\mathclose\bgroup\originalleft}
\newcommand{\rightR}{\aftergroup\egroup\originalright}

\DeclareMathOperator{\tr}{tr}

\newcommand{\LES}{\ell_\mathrm{ES}}
\newcommand{\LSD}{\ell_\mathrm{SD}}
\newcommand{\RHSD}{R_\mathrm{H}^\mathrm{SD}}
\newcommand{\RHES}{R_\mathrm{H}^\mathrm{ES}}
\newcommand{\RHESs}{R_\mathrm{H}^\mathrm{ES,s}}
\newcommand{\RHESl}{R_\mathrm{H}^\mathrm{ES,l}}

\newcommand{\code}[1]{{\fontfamily{qcr}\selectfont #1}}

\usepackage[usenames,dvipsnames]{xcolor}

\newcommand{\rSI}{SM}

\definecolor{myblue}{rgb}{0, 0, 0.901}
\definecolor{myorange}{rgb}{0.901, 0.364, 0.090}
\definecolor{mygreen}{rgb}{0.160, 0.6, 0.031}

\newcommand{\hypgeo}[2]{%
  \operatorname{%
    {\vphantom{\mathnormal{F}}}_{#1}%
    \kern-\scriptspace
    \mathnormal{F}_{#2}%
  }%
}

\newcommand{\bmr}{\bm{r}}

\newcommand{\bmu}{\bm{u}}
\newcommand{\bmv}{\bm{v}}
\newcommand{\bmw}{\bm{w}}

\newcommand{\bmF}{\bm{F}}

\newcommand{\bmI}{\bm{I}}

\newcommand{\bmR}{\bm{R}}

\newcommand{\bmT}{\bm{T}}

%% file: content.tex
\section{Introduction}

The aggregation of transmembrane and membrane-associated proteins is an important principle of membrane organization \cite{recouvreux2016molecular}. Membrane protein aggregation plays an important role in the recognition and transduction of cellular signals, but has also been proposed as a strategy to achieve high signaling reliability \cite{mugler2013spatial,case2019regulation,sanchez2023ligand,kim2024interplay,murai2024transmembrane}. Examples of aggregating transmembrane signaling proteins include Linker for Activation T-cells \cite{zhang2000association,houtman2006oligomerization,zeng2021plcgamma1}, glutamate receptors in the postsynaptic density \cite{macgillavry2013nanoscale,nair2013super,goncalves2020nanoscale}, Epidermal Growth Factor Receptor (EGFR) and Fibroblast Growth Factor Receptor 2 (FGFR2) \cite{lin2022receptor,lin2022two,phan2024grb2}. The latter two examples have been associated with cancer-related signaling \cite{sigismund2018emerging, krook2021fibroblast,lin2024emerging}, highlighting the importance of understanding the mechanisms of membrane protein aggregation. 

Membrane protein aggregation is also observed in other cellular processes. The crystalline glycoprotein coats that cover the inner cell surface of most prokaryotes (S-layers) assemble through the formation of smaller aggregates before crystallization \cite{chung2010self,sleytr2014s, sleytr2025s}. Some enveloped viruses, such as influenza, form their capsid with the help of transmembrane glycoproteins \cite{perlmutter2015mechanisms}. Similarly, clathrin assembles on cell membranes with the help of adapter proteins to induce clathrin-mediated endocytosis \cite{kaksonen2018mechanisms}. These examples further illustrate that membrane protein aggregation is an important mechanism of membrane organization. In this article, we seek to address the question of what determines the diffusivity of membrane protein aggregates. This fundamental question is not only relevant for understanding aggregation mechanisms but may also be relevant for their function. 

\begin{figure*}
    \centering
    \includegraphics[scale=1]{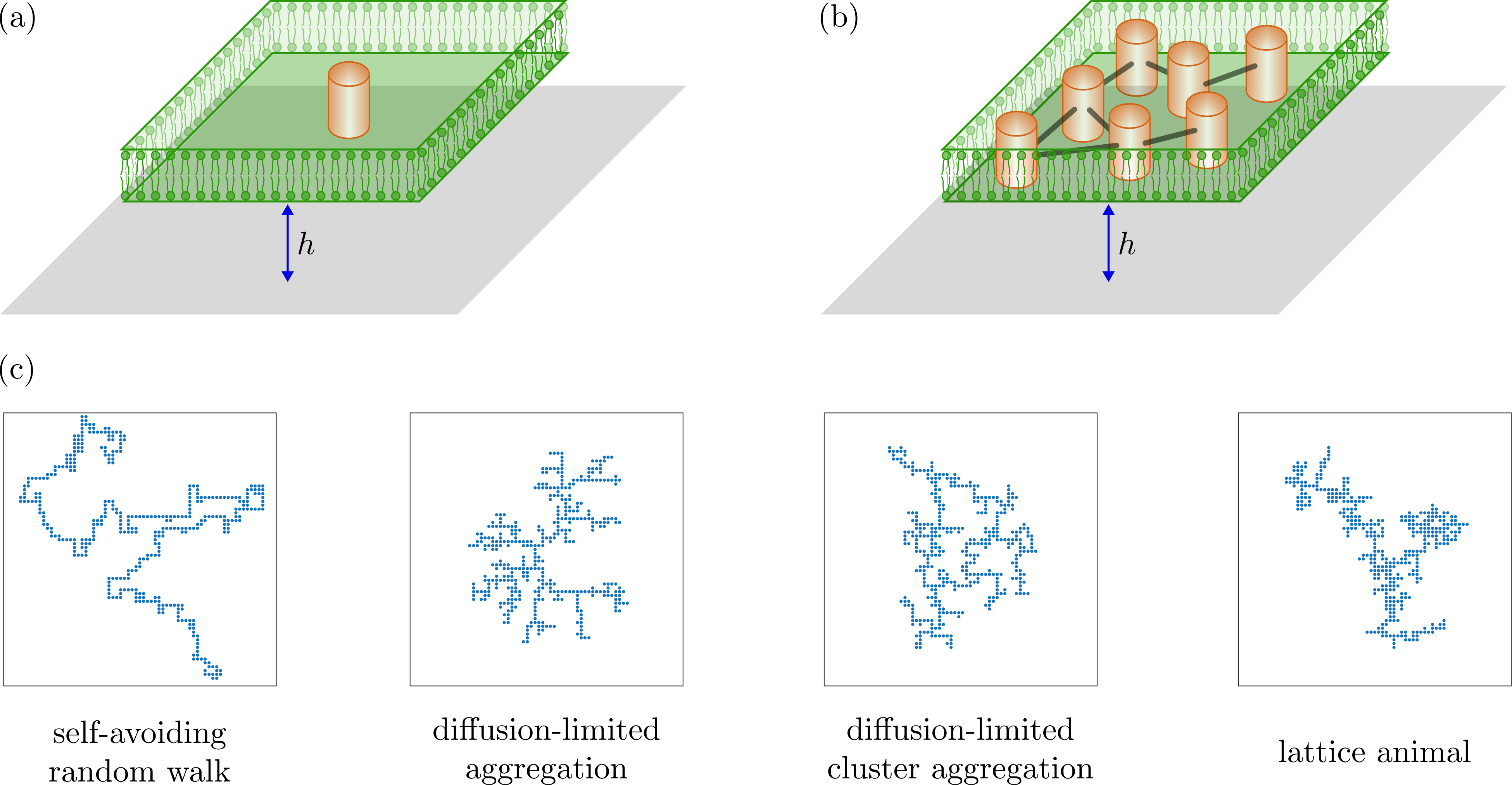}
    \caption{Setups for single particle diffusion (a) and the diffusion of a protein aggregate (b), here drawn as crosslinked, a distance $h$ away from a rigid substrate. We apply KR theory to the aggregate types in (c), where each realization of a self-avoiding random walk (SAW), diffusion-limited aggregation (DLA), diffusion-limited cluster aggregation (DLCA) has $400$ particles and the lattice animal (LA) has $400$ edges (not shown).}
    \label{fig:overview}
\end{figure*}

The diffusion of individual membrane proteins in lipid membranes was described in the pivotal work by Saffman and Delbr\"uck (SD) (Fig.~\ref{fig:overview}\,(a) with $h \rightarrow \infty$). The authors realized that finding the drag force on a particle in a membrane requires consideration of the surrounding fluid to avoid Stokes' paradox \cite{saffman1975brownian,saffman1976brownian}. Using the Stokes-Einstein relation, this led to an expression for the diffusivity of a cylindrical membrane inclusion, 
\begin{align}
D_\mathrm{SD} = \frac{k_\mathrm{B} T}{\xi_\mathrm{SD}} = \frac{k_\mathrm{B} T}{4 \pi \zeta} \left( \ln{\frac{2\ell_\mathrm{SD}}{a}} - \gamma \right)~, \label{eq:DSD}
\end{align}
where $\xi_\mathrm{SD}$ denotes the drag coefficient, $\zeta$ is the membrane viscosity, $\ell_\mathrm{SD} = \zeta/(\eta^+ + \eta^-) = \mathcal{O}\leftR(1\mu \mathrm{m}\rightR)$ \cite{faizi2022vesicle} is the SD length scale, $\eta^\pm$ are the bulk viscosities above and below the membrane, $a$ is the particle radius, and $\gamma$ is the Euler--Mascheroni constant. This result is limited to the regime where $a \ll  \ell_\mathrm{SD}$ and to the assumption that the membrane is free-standing, i.e. not interacting with a wall or another membrane. The first assumption was relaxed in Ref.~\cite{hughes1981translational} by Hughes, Pailthorpe, and White (HPW), which revealed a linear scaling $D_\mathrm{SD} \propto a^{-1}$ for large particles, $a \gg \ell_\mathrm{SD}$.

Evans and Sackmann (ES) \cite{evans1988translational} relaxed the second assumption and adapted SD theory to the case where the membrane is situated a small distance $h \ll \LSD$ away from a rigid wall or substrate, as is the case for supported lipid bilayers. (Fig.~\ref{fig:overview}\,(a)).  
This led to the diffusivity
\begin{align}
	D_\mathrm{ES} = \frac{k_\mathrm{B} T}{\xi_\mathrm{ES}} = \frac{k_\mathrm{B} T}{4 \pi \zeta} \left( \frac{1}{4} \varepsilon^2 + \varepsilon \frac{K_1(\varepsilon)}{K_0(\varepsilon)} \right)^{-1}~, \label{eq:DES}
\end{align}
where $\xi_\mathrm{ES}$ is the drag coefficient for the ES case, $K_i$ are modified Bessel functions of the second kind and $\varepsilon = a/\ell_\mathrm{ES}$ is the non-dimensional radius expressed in terms of the ES length scale, $\ell_\mathrm{ES} = \sqrt{\zeta/b_\mathrm{s}} \equiv \sqrt{h\zeta/\eta}$. In the last expression, the first equality uses the friction coefficient $b_\mathrm{s}$ to describe membrane-substrate interactions whereas the second equality is a result of solving Stokes equations in the thin fluid region between the membrane and substrate. Using the latter relation, we find $\LES \approx 141~\mathrm{nm}$ for $h=20~\mathrm{nm}$ and $\LES \approx 45~\mathrm{nm}$ for $h=2~\mathrm{nm}$ which is significantly shorter than the SD length scale.

In this article, we investigate the instantaneous translational diffusion coefficient of  aggregates of identical transmembrane proteins. To this end, we first discuss Kirkwood-Riseman theory in the context of the diffusion of proteins on lipid membranes and subsequently apply the theory to four different aggregation models. We then show that the translational diffusivity of all models is well characterized by appropriate hydrodynamic radii. Crucially, these newly-derived hydrodynamic radii exclusively depend on the interparticle distances, analogous to the three-dimensional case. Finally, we show that the aggregate diffusion coefficients can be accurately approximated using only the aggregate outline and number of particles in it. This approach is especially useful when the interparticle distances cannot be resolved.

\section{Theory}

We employ Kirkwood-Riseman theory (KR) \cite{kirkwood1948intrinsic} to evaluate the diffusion coefficients of aggregates of transmembrane proteins. KR theory is well-established in three dimensions 
\cite{edwards1984translational, meakin1985translational,lattuada2003hydrodynamic} and has also been applied to lipid membranes \cite{oppenheimer2009correlated,sokolov2018many, sorkin2021persistent}. It provides a self-consistent formalism that captures the instantaneous hydrodynamic interactions of a collection of $N$ particles. We begin by considering the in-plane force $\bmF_i$ that a particle $i$, located at $\bmr_i$, exerts on the membrane,
\begin{align}
    \bmF_i = -\xi \left( \bmv_i - \bmu_i \right)~, \label{eq:forceOnFluid}
\end{align}
where $\bmv_i$ is the velocity of the membrane at the center of particle $i$ if particle $i$ was not present, $\bmu_i$ is the velocity of particle $i$, and $\xi$ denotes the drag coefficient, defined by Eq.~\eqref{eq:forceOnFluid} itself. We further decompose the membrane velocity into a base flow, $\bmv_i^0$, and a perturbation due to the presence of other particles, $\bmv_i^\prime$, as $\bmv_i = \bmv_i^0 + \bmv_i^\prime$. Assuming Stokes flow, velocity perturbations are linearly related to the forces exerted by the other particles, 
\begin{align}
    \bmv_i^\prime = \sum_{\substack{j=1,~ i\neq j}}^N \bmT\leftR(\bmr_i, \bmr_j\rightR) \bmF_j~, \label{eq:veloPerturbation}
\end{align}
where $\bmT\leftR(\bmr_i, \bmr_j\rightR)$ is a $2\times 2$ tensor. We address the form of $\bmT\leftR(\bmr_i, \bmr_j\rightR)$ for lipid membranes below. Substituting Eq.~\eqref{eq:veloPerturbation} into Eq.~\eqref{eq:forceOnFluid} and introducing the shorthand notation $\bmT_{ij} = \bmT\leftR(\bmr_i, \bmr_j\rightR)$ leads to the self-consistent force balance
\begin{align}
    \bmF_i = -\xi \left( \bmv_i^0 - \bmu_i \right) - \xi \sum_{\substack{j=1,~ i\neq j}}^N \bmT_{ij} \bmF_j~.  \label{eq:forceBalanceKR}
\end{align}
It is convenient to choose the base and particle velocities to be identical for all particles, $\bmv_i^0 = \bmv^0$ and $\bmu_i = \bmu$, and define $\bmw = \bmv^0 - \bmu$. The aggregate drag coefficient tensor $\bm{\Xi}$ is then defined by summing over all forces, $\bmF = \sum_{i=1}^N \bmF_i$, and setting
\begin{align}
    \bmF = -\bm{\Xi} \bmw~. \label{eq:XiTensorDef}
\end{align}
We then define the scalar diffusion coefficient using the Stokes-Einstein relation as 
\begin{align}
    D =  \frac{k_\mathrm{B}T}{2}\tr{\leftR(\bm{\Xi}^{-1}\rightR)}~. \label{eq:scalarDDef}
\end{align}

Before proceeding, we note that Eq.~\eqref{eq:forceBalanceKR} describes a force balance on the membrane and not on the aggregate itself. Therefore, forces arising from direct interactions between particles do not appear here. This implies that KR theory can be applied to any instantaneous collection of particles and that our results are independent of the precise aggregation mechanism, which may, for example, be mediated by weak, non-covalent interactions, polymerization, or by lipid effects \cite{katira2016pre,case2019regulation}. Yet, we do assume that aggregates disperse and change their configurations more slowly than the diffusion timescale \cite{sorkin2021persistent}. However, in Sec.~\ref{sec:ria}, we show that the latter assumption can be relaxed.

Applying KR theory requires an expression for the tensor $\bmT_{ij}$. Under the assumption of point particles, this is simply the Oseen tensor. While the point-particle approximation is suitable at large particle distances, we consider the case where interparticle distance $r_{ij} = || \bmr_i - \bmr_j ||$ can be on the order of the particle size. In this case, the membrane-equivalent of the Rotne--Prager--Yamakawa (RPY) tensor provides a more suitable approximation because it accounts for the particles' finite size. For free-standing lipid membranes (SD limit), Sokolov and Diamant \cite{sokolov2018many} derived the RPY-type tensor using the variational approach by Rotne and Prager \cite{rotne1969variational}. In Secs.~1.2 and 2.2 of the Supplementary Material (\rSI), we derive the RPY-type tensor for both the SD and ES cases using Yamakawa's multipole expansion approach. In the SD limit, this yields the same result for small particle distances as in Ref.~\cite{sokolov2018many}. In Secs.~1 and 2.4 of the \rSI, we also show that the Oseen and RPY-type tensors are always positive definite for non-overlapping particles, thus guaranteeing positive dissipation rates \cite{rotne1969variational, wajnryb2013generalization,zuk2014rotne, sokolov2018many}. The expressions for the RPY-type tensor in the SD and ES cases, denoted by $\bmT_{ij}^\mathrm{SD}$ and $\bmT_{ij}^\mathrm{ES}$, respectively, are summarized in Sec.~\ref{app:Tij}.

\section{Results from Kirkwood-Riseman Theory}
We apply KR theory to four different types of aggregates composed of identical particles with radius $5~\mathrm{nm}$ arranged on a square lattice with spacing $15~\mathrm{nm}$. Specifically, we consider self-avoiding random walks (SAW), lattice animals (LA), and aggregates formed through diffusion-limited aggregation (DLA) and diffusion-limit cluster aggregation (DLCA). Examples of realizations of each of the aggregates are shown in Fig.~\ref{fig:overview}\,(c).  For our results, we generate aggregates with up to $1000$ particles (SAW, DLA, DLCA) or edges (LA). We then compute the diffusion coefficient of each aggregate using Eqs.~\eqref{eq:forceBalanceKR}--\eqref{eq:scalarDDef} (see Sec.~\ref{app:simDeets} for details). 

\begin{figure}
    \centering
    \includegraphics[scale=1]{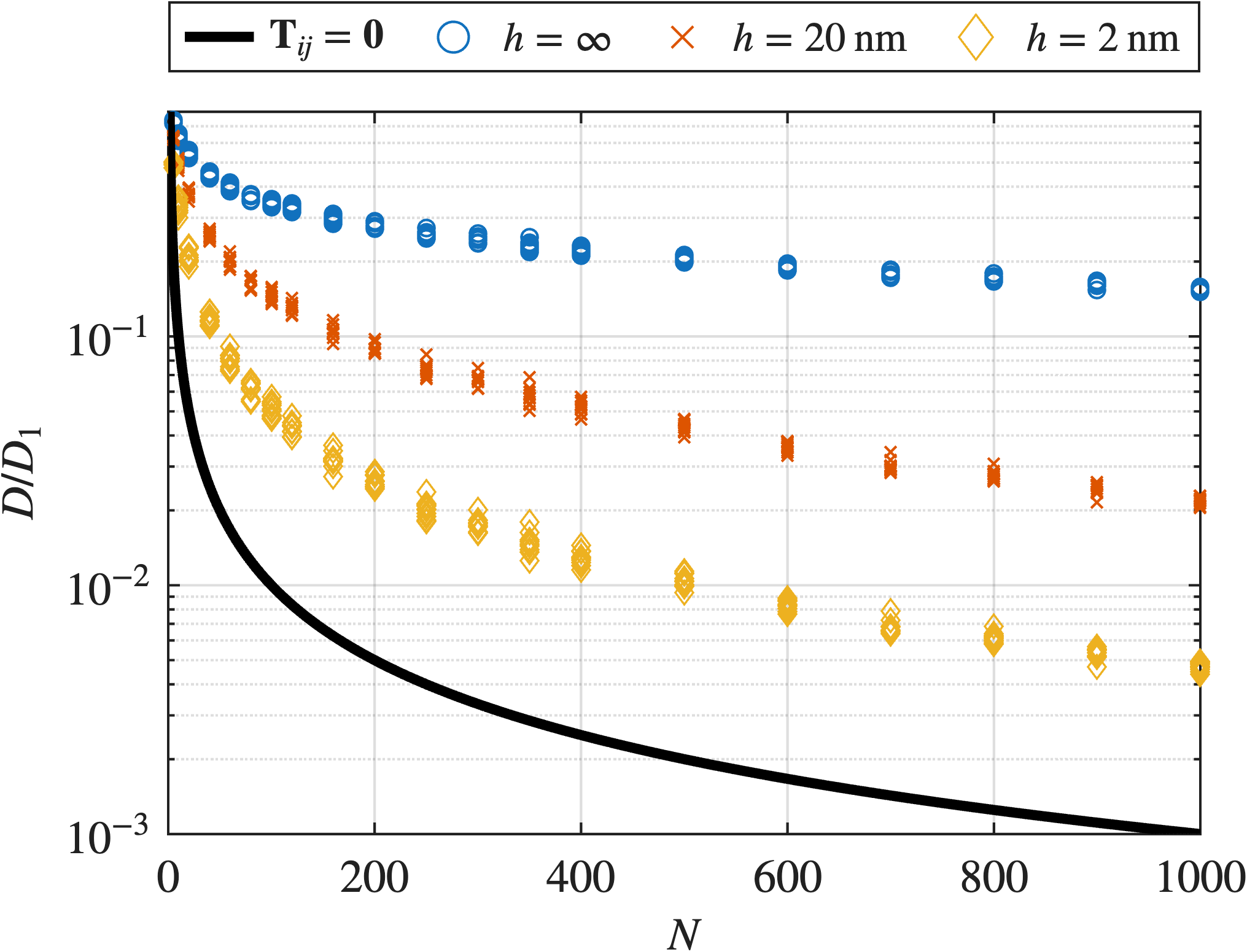}
    \caption{Normalized diffusion coefficients for diffusion-limited aggregation (DLA) with a varying number of particles $N$ at different wall distances $h$. Each marker corresponds to a separate realization of DLA (see Sec.~\ref{app:simDeets} for simulation details). We find that the diffusion coefficients decay more rapidly with decreasing distances to the wall. Comparison to the case without hydrodynamic interactions between the particles (black full line, $\bmT_{ij} = \bm0$) shows a difference of up to two orders of magnitude in the diffusion coefficients, illustrating the importance of accounting for hydrodynamic interactions.}
    \label{fig:Dplot_DLA_hvary}
\end{figure}

Figure~\ref{fig:Dplot_DLA_hvary} shows the normalized diffusion coefficients of aggregates formed through DLA as a function of the number of particles in each aggregate for different wall distances $h$. Each marker corresponds to one realization of an aggregate. We also plot the free-draining limit ($\bmT_{ij} = \bm0$), where hydrodynamic interactions are neglected, leading to the scaling $D \propto N^{-1}$. In comparison, accounting for hydrodynamic interactions through KR theory yields significantly larger diffusivities for all wall distances. This is a result of hydrodynamic screening \cite{witten2010structured}: The flow perturbation caused by an individual particle also induces perturbations in its surrounding, resulting in an effectively lower drag compared to the free-draining limit. Due to their highly viscous nature, hydrodynamic screening has a significant effect in lipid membranes. However, we also observe that these effects decrease with decreasing wall distance $h$. This can be explained by the characteristic length $\LES$ over which hydrodynamic perturbations decay in the ES limit (cf. Eq.~\eqref{eq:ES_Tij_finite}), which scales as $\LES \propto \sqrt{h}$. This scaling results from increasing velocity gradients between the membrane and wall with decreasing wall distance, leading to an increased drag on the membrane and a faster decay of flow perturbations. The reduced screening lowers the diffusivity such that it approaches the free-draining limit with decreasing wall distance. 

\begin{figure*}
    \centering
    \includegraphics[scale=1]{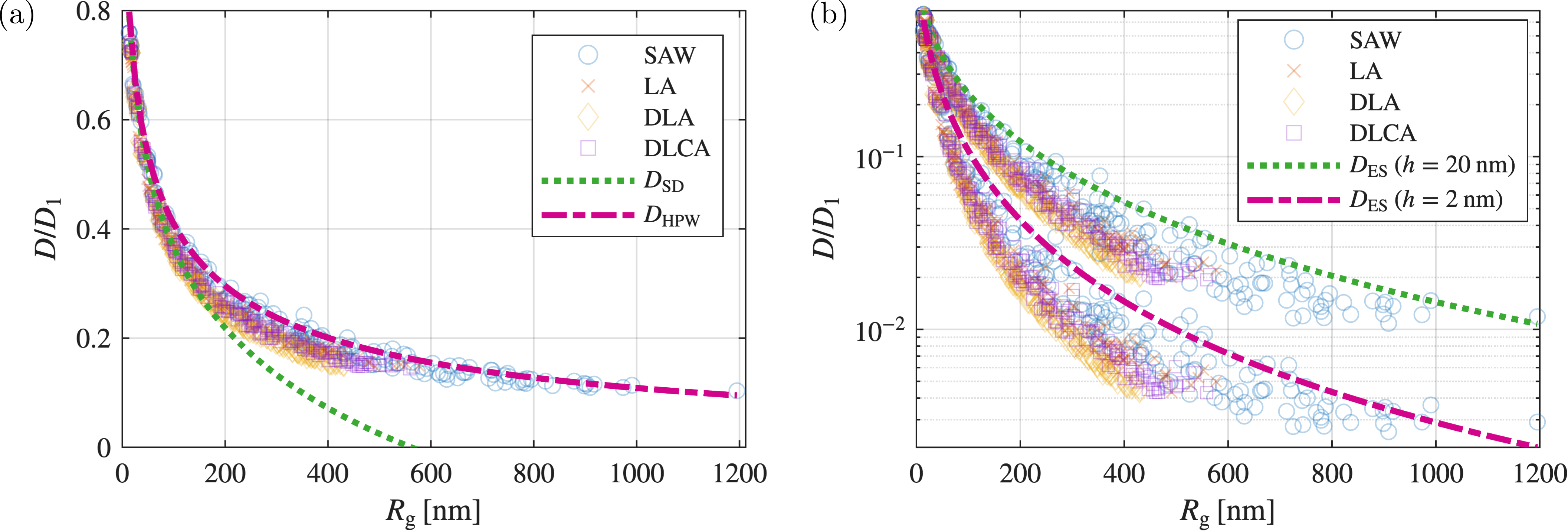}
    \caption{Diffusion coefficients for each realization of all aggregation types plotted against the radius of gyration defined in Eq.~\eqref{eq:Rgdef} for the SD case in (a) and for the ES case with $h=20~\mathrm{nm}$ and $h=2~\mathrm{nm}$ in (b). In the SD case, $R_\mathrm{g}$ groups the data well with a relatively small spread in the data. However, replacing the particle radius by $R_\mathrm{g}$ in Eq.~\eqref{eq:DSD} ($D_\mathrm{SD}$) yields a poor approximation for large aggregates. In contrast, the fit by Petrov and Schwille \cite{petrov2008translational} of HPW theory ($D_\mathrm{HPW}$), valid for large particles, provides a good approximation of the data. In the ES case, we observe a significant spread in the data and substituting $R_\mathrm{g}$ in Eq.~\eqref{eq:DES} ($D_\mathrm{ES}$) does not approximate the results well.}
    \label{fig:SD_ES_Rg}
\end{figure*}

To relate the aggregate diffusion coefficients to the aggregates' size, Fig.~\ref{fig:SD_ES_Rg} shows the diffusion coefficients plotted against the radius of gyration $R_\mathrm{g}$ for all aggregate types, where 
\begin{align}
    R_\mathrm{g} =\sqrt{\frac{1}{N} \sum_{i=1}^N \left( \bmr_i - \bmR_0 \right)^2}~, \label{eq:Rgdef}
\end{align}
with $\bmR_0$ denoting the aggregate center of mass. While the diffusivities only show a moderate spread in the SD case across different aggregate types (Fig.~\ref{fig:SD_ES_Rg}\,(a)), they vary significantly in the ES case (Fig.~\ref{fig:SD_ES_Rg}\,(b)). We further approximate the aggregates by single solid particles by replacing the particle radius by $R_\mathrm{g}$ in Eqs.~\eqref{eq:DSD} and \eqref{eq:DES}. 
In the SD limit, this provides a good approximation of the KR results for small aggregates but decays too rapidly for large aggregates. Since SD theory assumes the particle radius is much smaller than the SD length scale, we can, in fact, not expect SD theory to reproduce the correct qualitative behavior for large aggregates. Therefore, we also compare the SD results to HPW theory \cite{hughes1981translational}, which captures the correct trend. In the ES case, Eq.~\eqref{eq:DES} provides a good approximation for small aggregates but systematically predicts diffusivities larger than obtained by KR theory for large aggregates.

\section{Hydrodynamic Radii}
In the preceding section, we found that the radius of gyration may not accurately predict the diffusion coefficients of protein aggregates. In the following, we derive approximate hydrodynamic radii, following Refs.~\cite{bloomfield1967frictional,de1977hydrodynamic}, which serve as better proxies of aggregate diffusivity. To this end, we assume that the forces on the right-hand side of Eq.~\eqref{eq:forceBalanceKR} can be approximated by $\bmF_j \approx - \xi \bmw$. Summing Eq.~\eqref{eq:forceBalanceKR} over all particles leads to (cf. Eq.~\eqref{eq:XiTensorDef}; see Sec.~3 of \rSI\ for details)
\begin{align}
    \bm{\Xi} \approx N \xi\left( \bmI_\mathrm{s} - \frac{\xi}{N} \sum_{i=1}^N\sum_{j=1,~ j\neq i}^N \bmT_{ij}\right)~,
\end{align}
where $\bmI_\mathrm{s}$ is the $2\times 2$ identity tensor. We then define the hydrodynamic radii for the SD and ES cases, $\RHSD$ and $\RHES$, such that the diffusion coefficient is approximated by the respective single particle theories. In the SD case, we therefore set
\begin{align}
    \frac{k_\mathrm{B} T}{4 \pi \zeta} \left( \ln{\frac{2\ell_\mathrm{SD}}{\RHSD}} - \gamma \right) &= \frac{k_\mathrm{B} T}{2}  \tr\leftR(\bm{\Xi}^{-1}\rightR)~. \label{eq:RHSD_def}
\end{align}
By solving Eq.~\eqref{eq:RHSD_def} for $\RHSD$, we find (see Sec.~3.1 of \rSI)
\begin{widetext}
\begin{align}
    \RHSD = 2\LSD\left(\frac{a}{2\LSD}\right)^{1/N} \operatorname{exp}\leftR( -\frac{\pi}{2N^2} \sum_{i=1}^N \sum_{\substack{j=1 \\ j\neq i}}^N \left( H_0\leftR(\frac{r_{ij}}{\LSD}\rightR) - Y_0\leftR(\frac{r_{ij}}{\LSD}\rightR)\right) -  \frac{\gamma\left(N-1\right)}{N}\rightR)~, \label{eq:RHSD}
\end{align}
\end{widetext}
where $H_0$ and $Y_0$ denote the zeroth-order Struve functions and Bessel functions of the second kind, respectively. Substituting this result back into Eq.~\eqref{eq:RHSD_def} shows that the term $ \left(\frac{a}{2\LSD}\right)^{1/N} \operatorname{exp}\leftR(-  \frac{\gamma\left(N-1\right)}{N}\rightR)$ describes the contribution due to $N$ uncoupled particles, i.e. the free-draining limit, and the remaining terms describe the hydrodynamic corrections. We note here that substituting Eq.~\eqref{eq:RHSD} into Eq.~\eqref{eq:RHSD_def} and expanding the Struve and Bessel functions for small particle distances yields the instantaneous center of mass diffusion of a dispersing collection of particles obtained in Ref.~\cite{sorkin2021persistent}.

\begin{figure*}
    \centering
    \includegraphics[scale=1]{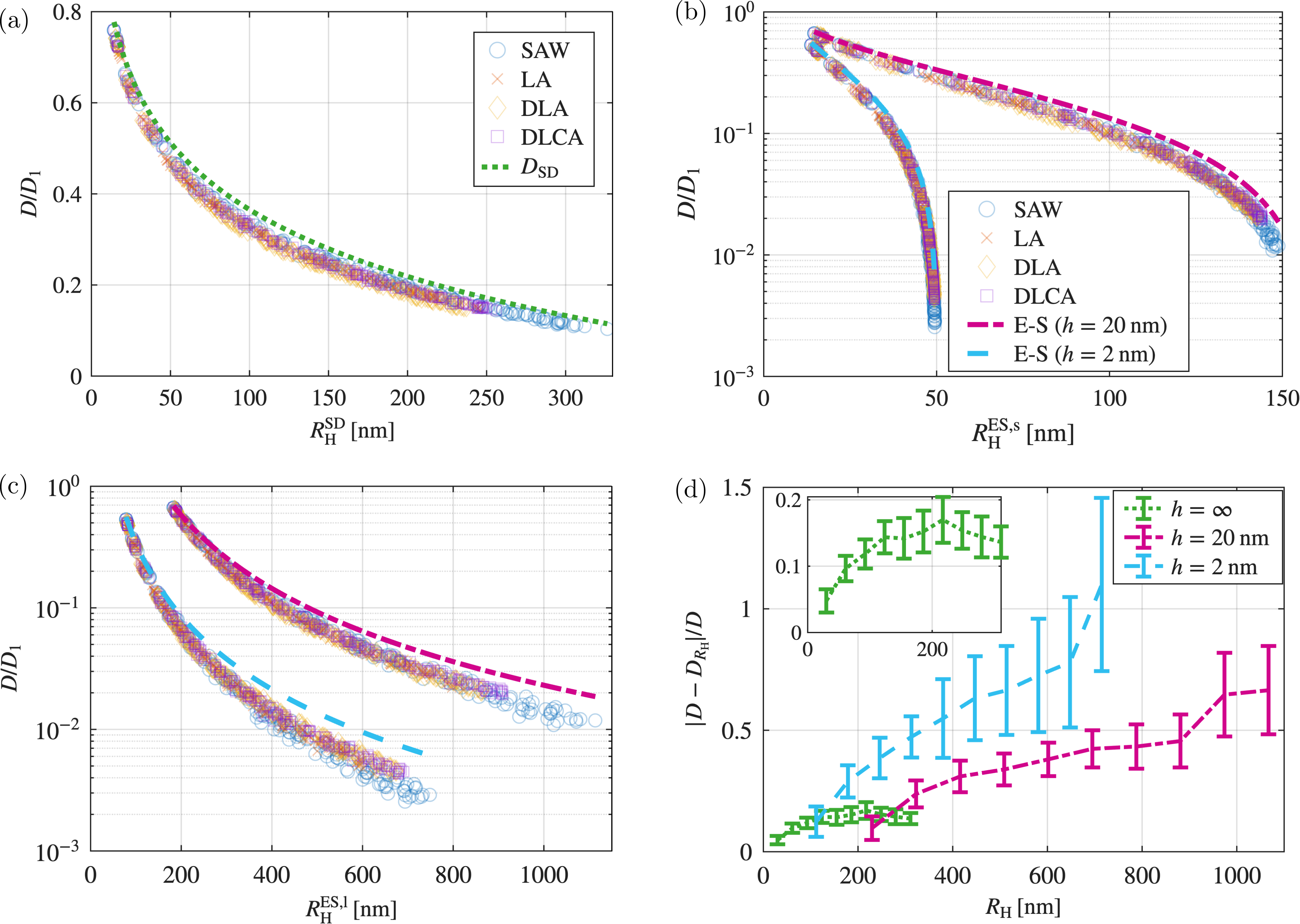}
    \caption{Diffusion coefficients plotted against the hydrodynamic radius for the SD case in (a) and for the small and large radius expansion of the ES case in (b) and (c), respectively. In all cases, we find that the data collapse well onto single curves. The lines in (a)--(c) correspond to the left-hand sides of Eqs.~\eqref{eq:RHSD_def}, \eqref{eq:RHESs} and \eqref{eq:RHESl}. Subfigure (d) shows the mean and standard deviation of the errors corresponding to these curves, binned by $\RHSD$ and $\RHESl$ across all aggregate types. While the error remains generally below $20\%$ for the SD case, it becomes exceedingly large for large aggregate in the ES case. Note that the legend in (b) also applies to (c).}
    \label{fig:SD_ES_RH}
\end{figure*}

While Eq.~\eqref{eq:RHSD_def} can be readily solved for $\RHSD$, an analogous definition in the ES case cannot be analytically inverted. Therefore, we expand Eq.~\eqref{eq:DES} for small and large hydrodynamic radii to obtain the asymptotic forms of $\RHES$, which we denote by $\RHESs$ and $\RHESl$, respectively, yielding
\begin{align}
    \frac{k_\mathrm{B}T}{4\pi\zeta}\left( \ln{\frac{2\LES}{\RHESs}}-\gamma\right) &= \frac{k_\mathrm{B} T}{2}  \tr\leftR(\bm{\Xi}^{-1}\rightR) ~, \label{eq:DESs}\\
    \frac{k_\mathrm{B}T}{\pi\zeta} \left(\frac{\LES}{\RHESl}\right)^2 &= \frac{k_\mathrm{B} T}{2}  \tr\leftR(\bm{\Xi}^{-1}\rightR)~.\label{eq:DESl}
\end{align}
This gives rise to the hydrodynamic radii (see Sec.~3.2 of \rSI)
\begin{widetext}
\begin{align}
    \RHESs &= 2\LES\left(\frac{a}{2\LES}\right)^{1/N} \operatorname{exp}\leftR( -\frac{1}{N^2} \sum_{i=1}^N \sum_{\substack{j=1 \\ j\neq i}}^N K_0\leftR(\frac{r_{ij}}{\LES}\rightR) - \frac{\gamma\left(N-1\right)}{N}\rightR)~, \label{eq:RHESs} \\[6pt]
    \RHESl &= \frac{2\LES }{\sqrt{\frac{4\pi\zeta}{N \xi_\mathrm{ES}} + \left(1-\frac{a^2}{2\LES^2}\right) \frac{1}{N^2} \sum_{i=1}^N \sum_{\substack{j=1,~j\neq i}}^N K_0\leftR( \frac{r_{ij}}{\LES} \rightR) }}~, \label{eq:RHESl}
\end{align}
\end{widetext}
where $K_0$ denotes the zeroth-order modified Bessel function of the second kind. Similar to the SD case, the term $2\LES\left(\frac{a}{2\LES}\right)^{1/N} \operatorname{exp}\leftR(-\frac{\gamma\left(N-1\right)}{N}\rightR)$ in Eq.~\eqref{eq:RHESs} and the first term under the square root in Eq.~\eqref{eq:RHESl} describe the hydrodynamic radii in the free-draining limit and the remaining terms capture the hydrodynamic interactions. It is important to note that the hydrodynamic radii are fully determined by the interparticle distances, just as in three dimensions but with a different functional form \cite{bloomfield1967frictional}. This is expected as the interparticle distances encode all information about the aggregate structure.

Figures \ref{fig:SD_ES_RH}(a)–(c) show the aggregate diffusion coefficients plotted against the hydrodynamic radii defined by Eq.~\eqref{eq:RHSD} (SD) and by Eqs.~\eqref{eq:RHESs}–\eqref{eq:RHESl} (ES). In all cases, the hydrodynamic radius produces a clear collapse of the KR results, irrespective of the underlying aggregate type. This demonstrates that the hydrodynamic radii capture the effective length scales describing the translational diffusion of the aggregates.

Comparing the KR data to the left-hand sides of Eqs.~\eqref{eq:RHSD_def}, \eqref{eq:DESs} and \eqref{eq:DESl} (dotted, dashed-dotted, and dashed lines in Figs.~\ref{fig:SD_ES_RH}\,(a)--(c)) shows that the hydrodynamic radii capture the qualitative behavior well but tend to overestimate the diffusion coefficients compared to KR theory. The errors, binned by the hydrodynamic radii $\RHSD$ and $\RHESl$, are plotted in Fig.~\ref{fig:SD_ES_RH}\,(d). For all wall distances, the error increases with increasing aggregate size. While the error remains below $20~\%$ in the SD case, it becomes exceedingly large for large aggregates in the ES case. We note here, however, that better approximations can be obtained by fitting the collapsed data. However, since the results depend on the system parameters such as the particle size, a fit would not be universal and should be obtained for specific aggregates under investigation. 

Finally, we note that the small and large radii expansions in Eqs.~\eqref{eq:RHESs} and \eqref{eq:RHESl} provide equivalent parameterizations of the diffusion coefficient. Therefore, both yield the same errors. However, Fig.~\eqref{fig:SD_ES_RH}\,(a) shows that $\RHESs$ approaches $\LES$ for large aggregates. In this case, the use of $\RHESl$ gives rise to more distinct radii. In contrast, the use of $\RHESl$ leads to less distinct radii for small aggregates.

Next, we assume that the particle distances are small, $r_{ij}/\ell \ll 1$, where $\ell$ describes either the SD or ES length scale. This allows us to express the hydrodynamic coupling terms in the hydrodynamic radii as 
\begin{align}   
    &
    \begin{rcases}
    \frac{\pi}{2N^2} \sum_{i,j\neq i} \left( H_0\leftR(\frac{r_{ij}}{\LSD}\rightR) - Y_0\leftR(\frac{r_{ij}}{\LSD}\rightR)\right) \\
    \frac{1}{N^2} \sum_{i,j\neq i} K_0\leftR(\frac{r_{ij}}{\LES}\rightR)
    \end{rcases}\approx \nonumber\\
    & \hspace{0.3\linewidth}\frac{1}{N^2}\sum_{i,j\neq i} \ln \frac{2\ell}{r_{ij}} - \frac{\gamma\left(N-1\right)}{N}~.
\end{align}
The first term on the right-hand side is related to the geometric mean ($\mathrm{GM}$) by
\begin{align}
    \frac{1}{N^2}\sum_{i,j\neq i} \ln \frac{2\ell}{r_{ij}} = \frac{N-1}{N} \, \ln\leftR(\mathrm{GM}\leftR(\frac{2\ell}{r_{ij}}\rightR)\rightR)~, \label{eq:hydroGMExpression}
\end{align}
where we again only consider $i \neq j$. Thus, the hydrodynamic radii indicate that hydrodynamic coupling affects aggregate diffusion in terms of the logarithm of the geometric mean of the inverse interparticle distances. 

\section{Random-interior approximation} \label{sec:ria}
In the previous section, we showed that the aggregate diffusivity is well-characterized by the appropriate hydrodynamic radii. However, as shown in Fig.~\ref{fig:SD_ES_RH} (d), this may introduce significant errors. In addition, evaluating the hydrodynamic radii requires the interparticle distances, which may not be readily available from experimental measurements \cite{ober2004localization,abouelkheir2024investigations,sun2025cryo}. In the following, we show that we can circumvent these two challenges by using approximate representations of the aggregates. 

Equation~\eqref{eq:hydroGMExpression} indicates that, to lowest order, the geometric mean and not the precise distribution of particle distances characterizes hydrodynamic effects. This motivates us to approximate the aggregate by an equivalent structure. Specifically, suppose we have a representation of the outline of the aggregate, discussed in more detail below. If this outline is a good representation of the original aggregate, we may randomly distribute the particles inside the outline, hoping to recover a similar geometric mean of particle distances. We refer to this method as the \textit{random-interior approximation}.

\begin{figure}
    \centering
    \includegraphics[scale=1]{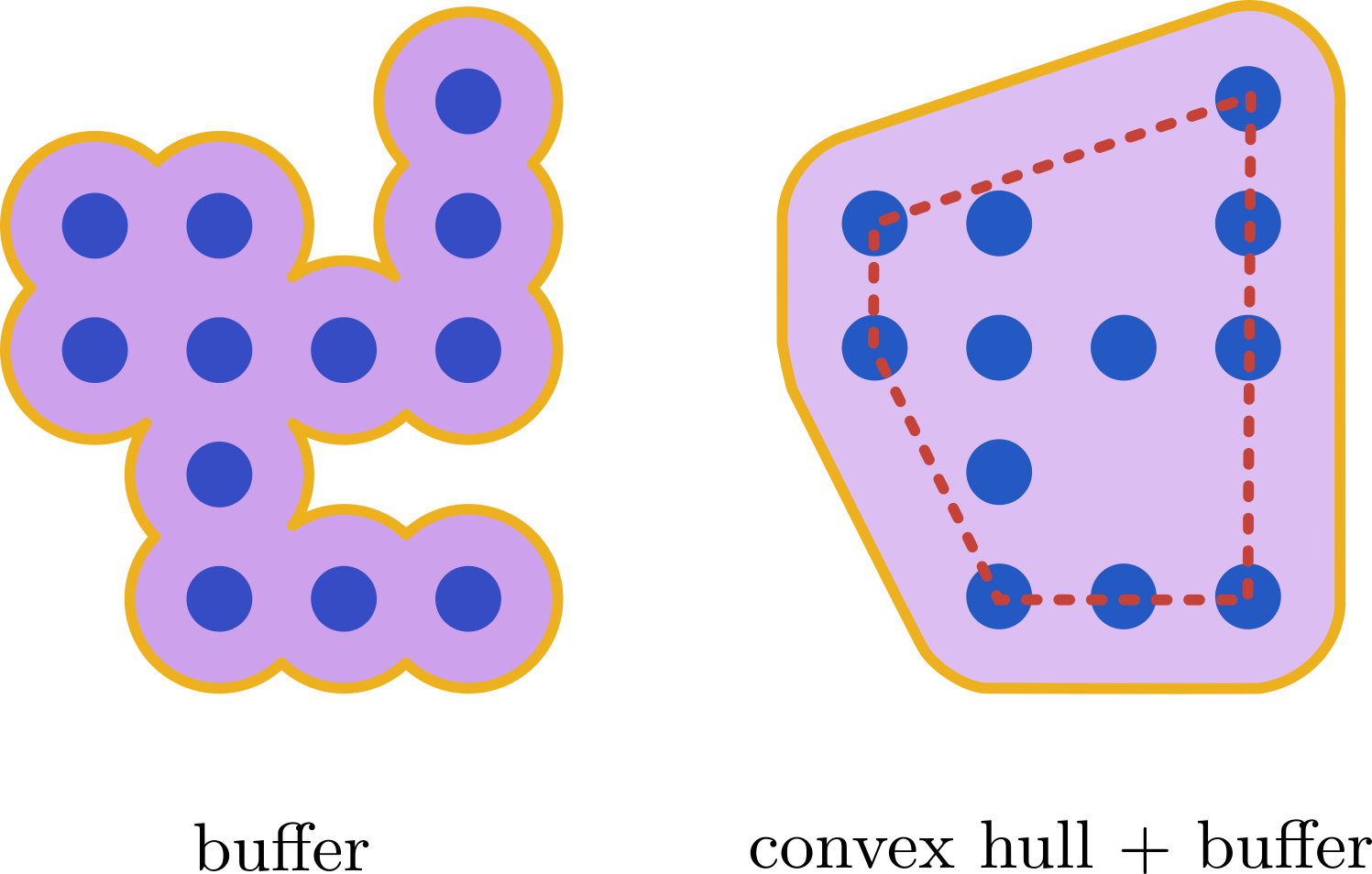}
    \caption{Outlines used to approximate an aggregate (blue circles). The buffer (left) describes all points that are within a radius of the original geometry (purple area). Alternatively, we can describe the aggregate by its convex hull (red dashed line) and a buffer around it (purple area).}
    \label{fig:outline_schematic}
\end{figure}

\begin{figure*}
    \centering
    \includegraphics[scale=1]{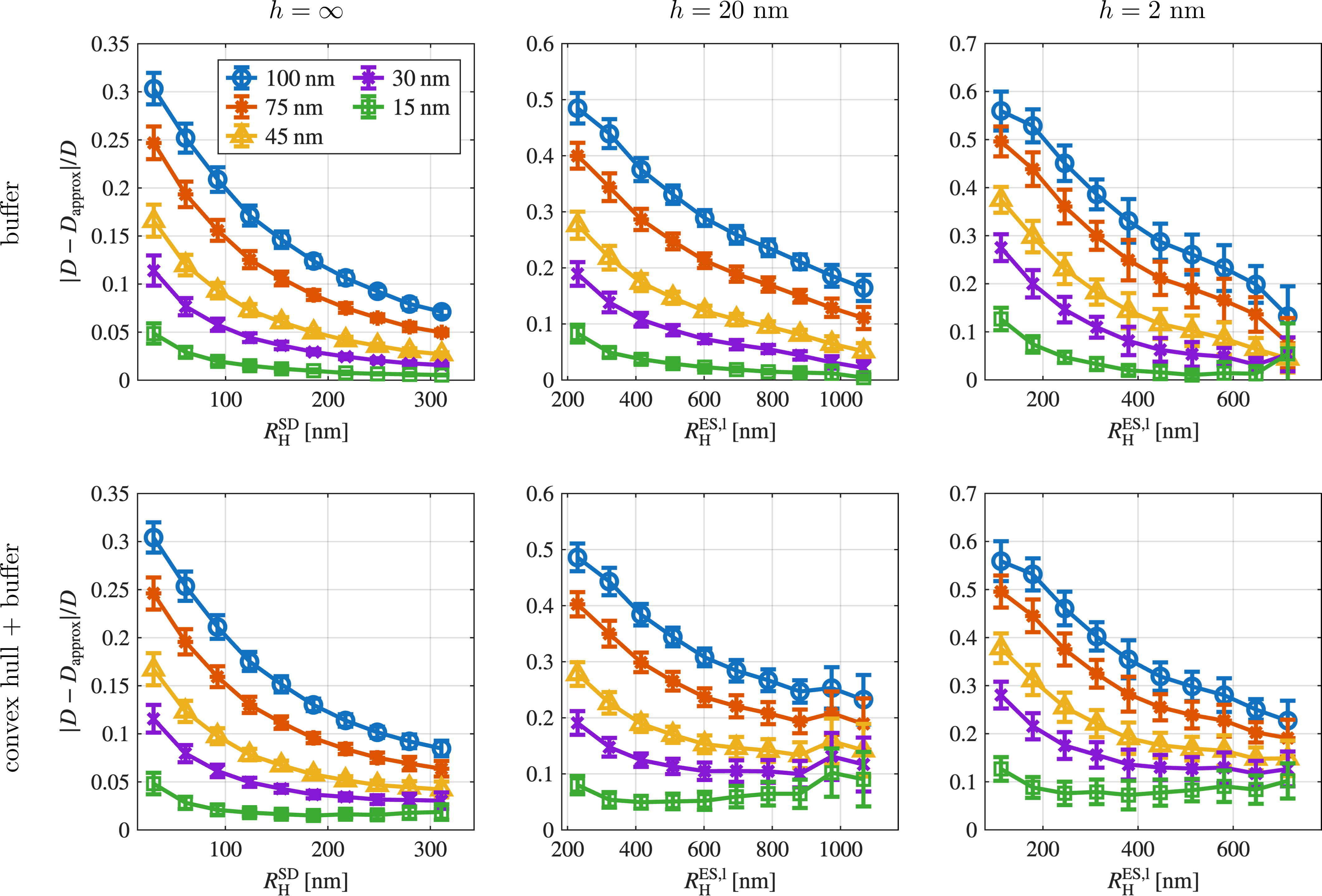}
    \caption{Error resulting from the buffer (top row) and convex hull (bottom row) approximations for $h=\infty$, $h=20~\mathrm{nm}$, and $h=2~\mathrm{nm}$ (left to right). For every aggregate realization, we generated $10$ samples in the respective aggregate outline. The errors were binned by the hydrodynamic radii $\RHSD$ and $\RHESl$ for the SD and ES case, respectively. The markers indicate the mean and the errorbars indicate the standard deviation. In all cases, the error decreases with decreasing $L_\mathrm{max}$ and increases with decreasing wall distance $h$.}
    \label{fig:ApproxError}
\end{figure*}

The proposed approach raises the question of how to choose an appropriate outline. During imaging, the position of a particle in the membrane can only be determined with limited accuracy \cite{ober2004localization}. We model the uncertainty of a particle's position by a circular shell with specified radius $L_\mathrm{max}$ and define the outline of an aggregate as the union of the shells of all particles (Fig.~\ref{fig:outline_schematic}, left). This construction is known as a buffer, defined by the set of points that are less than a specified distance $L_\mathrm{max}$ from the original geometry. Therefore, we simply refer to this first method as the buffer outline. The second method is motivated by an observation in three dimensions. Here, it has been shown that the mobility of a collection of particles can be approximated by its convex hull \cite{fleming2018hullrad}. We amend this approach by considering the buffer of the convex hull to account for the limited accuracy of the particle positions (Fig.~\ref{fig:outline_schematic}, right).

The implementation details of the random-interior approximation are provided in Sec.~\ref{app:simDeets}. The resulting errors compared to the exact KR are shown in Fig.~\ref{fig:ApproxError} for different values of $L_\mathrm{max}$ shown in the legend. We again binned the error by the hydrodynamic radii across all aggregate types and plotted its mean and standard deviation. The buffer method is shown in the top row. For all wall distances, we find that the error for the outline with the tightest fit ($L_\mathrm{max} = 15~\mathrm{nm}$) yields errors below $10\%$, confirming that the exact particle locations are not required to obtain an accurate approximation of the aggregate diffusivity. Increasing $L_\mathrm{max}$, and therefore capturing the outline of the aggregate less accurately, increases the error. In contrast, the error decreases with increasing aggregate size. This is expected as the fraction of additional area introduced by the buffer reduces. We further find that the error increases with a decreasing distance from the wall. This is due to the decreasing characteristic length scale of velocity perturbations ($\LSD$, $\LES$), making the random-interior approximation more sensitive to deviations in the interparticle distances. Finally, these findings suggest that the aggregate diffusion coefficients are insensitive to changes in the internal configuration of the aggregate. Therefore, our results remain valid over timescales longer than the timescale associated with these changes as long as the aggregate outline remains unaffected. 

The convex hull approximation is shown in the bottom row of Fig.~\ref{fig:ApproxError}. Here, we find that the errors are comparable to the buffer method in all cases. Only for the smaller values of $L_\mathrm{max}$ do we find slightly larger errors with the convex hull approach. Yet, the trends described in the previous paragraph are similarly found here. Therefore, we can conclude that determining the convex hull of the aggregate with adequate accuracy is sufficient to obtain a satisfactory approximation of the aggregate diffusivity. This result is highly relevant to experimental measurements where the precise particle locations are often not known but the convex hull can be estimated.

\section{Conclusion}
In this article, we described how hydrodynamics affect the diffusivity of transmembrane protein aggregates. Specifically, we reported the length scales characterizing aggregate diffusion and showed how the diffusivity can be approximated by the aggregate outline. These results provide a theoretical foundation for studying the formation and function of membrane protein aggregation as well as a practical method that can be invoked for experimental measurements. Nonetheless, it is well-understood that membranes should not simply be viewed as stationary, two-dimensional sheets. Thus, it is important to explore how factors such as membrane deformations \cite{reynwar2007aggregation,stachowiak2012membrane,sahu2017irreversible}, finite thickness effects \cite{omar20252+, lipel2025finite}, and lipid composition \cite{katira2016pre, wang2023coupling} affect the results presented here.  

\section{Methods}

\subsection{Rotne--Prager--Yamakawa-type tensors} \label{app:Tij}
The form of the RPY-type tensor differs between the SD and ES cases. In Sec.~1.2 of the \rSI, we show that for the SD case, it is given by
\begin{widetext}
\begin{align}
    \bmT_{ij}^\mathrm{SD}\leftR(\bmr_{ij}\rightR) = & \frac{1}{4\zeta} \left[ \left( \vphantom{\frac{a^2 H_{-1}\leftR(\frac{r_{ij}}{\ell_\mathrm{SD}}\rightR)}{r_{ij}\ell_\mathrm{SD}^2} }  H_0\leftR(\frac{r_{ij}}{\ell_\mathrm{SD}}\rightR)\left(1-\alpha\right) - \frac{\alpha \LSD}{r_{ij}}H_{-1}\leftR(\frac{r_{ij}}{\ell_\mathrm{SD}}\rightR) -\frac{\ell_\mathrm{SD}}{r_{ij}}H_1\leftR(\frac{r_{ij}}{\ell_\mathrm{SD}}\rightR)  \right. \right. \nonumber \\
    & \hspace{0.2\linewidth}\left.\left. - \frac{1-\alpha}{2}\left(Y_0\leftR(\frac{r_{ij}}{\ell_\mathrm{SD}}\rightR) - Y_2\leftR(\frac{r_{ij}}{\ell_\mathrm{SD}}\rightR)\right) +\frac{2\ell_\mathrm{SD}}{\pi r_{ij}}\left( \frac{\ell_\mathrm{SD}}{r_{ij}} + \alpha\right) \right) \bmI_\mathrm{s} \right. \nonumber \\
    &\left. -\left( H_0\leftR(\frac{r_{ij}}{\ell_\mathrm{SD}}\rightR)\left( 1 - \alpha\right) - \frac{2\alpha\LSD}{r_{ij}} H_{-1}\leftR(\frac{r_{ij}}{\ell_\mathrm{SD}}\rightR) -\frac{2\ell_\mathrm{SD}}{r_{ij}}H_1\leftR(\frac{r_{ij}}{\ell_\mathrm{SD}}\rightR) \right.\right.\nonumber \\
    &\hspace{0.2\linewidth} \left.\left.+ Y_2\leftR(\frac{r_{ij}}{\ell_\mathrm{SD}}\rightR)\left( 1 - \alpha\right)  +\frac{\ell_\mathrm{SD}}{\pi r_{ij}} \left( \frac{4\ell_\mathrm{SD}}{r_{ij}} + 2\alpha \right) \vphantom{\frac{a^2 H_{-1}\leftR(\frac{r_{ij}}{\ell_\mathrm{SD}}\rightR)}{r_{ij}\ell_\mathrm{SD}^2} } \right) \frac{\bmr_{ij} \otimes \bmr_{ij}}{r_{ij}^2} \right]~, \label{eq:SD_RPY_full}
\end{align}
where we defined $\alpha = \frac{a^2}{2\ell_\mathrm{SD}^2}$ and used the Struve functions $H_i$, Bessel functions of the second kind $Y_i$, and the particle distance vector $\bmr_{ij} = \bmr_i - \bmr_j$ with norm $r_{ij} = || \bmr_{ij} ||$. For a vanishing particle radius $a$, we recover the form used in Ref.~\cite{oppenheimer2009correlated}. For the ES case, we find (see Sec.~2.2 of \rSI)
\begin{align}
    \bmT_{ij}^\mathrm{ES} &= \frac{1}{2\pi \zeta} \left\{ \left[ \left( K_0\leftR(\frac{r}{\ell_\mathrm{ES}}\rightR) + \frac{\ell_\mathrm{ES}}{r} K_1\leftR(\frac{r}{\ell_\mathrm{ES}}\rightR)\right) \left( 1 + \frac{a^2}{2\ell_\mathrm{ES}^2}\right) - \frac{\ell_\mathrm{ES}^2}{r^2}\right]\bmI_\mathrm{s} \right.\nonumber\\
    &\hspace{3cm}\left. + \left[ \frac{2\ell_\mathrm{ES}^2}{r^2} - \left(K_0\leftR(\frac{r}{\ell_\mathrm{ES}}\rightR) + \frac{2\ell_\mathrm{ES}}{r}K_1\leftR(\frac{r}{\ell_\mathrm{ES}}\rightR) \right)\left( 1 + \frac{a^2}{2\ell_\mathrm{ES}^2}\right)  \right] \frac{\bmr \otimes \bmr}{r^2} \right\}~. \label{eq:ES_Tij_finite}
\end{align}
where $K_i$ are the modified Bessel functions of the second kind. In the limit of the particle radius $a$ vanishing, Eq.~\eqref{eq:ES_Tij_finite} reduces to the result derived in Ref.~\cite{ramachandran2011dynamics} (see Sec.~2 of \rSI). 

\end{widetext}

\subsection{Simulation details} \label{app:simDeets}
\paragraph{Aggregation models:} SAWs are implemented using the Pivot algorithm \cite{madras1988pivot}; LAs are generated using \textit{Algorithm S} of Ref.~\cite{van1997metropolis}; DLA follows the approach originally described by Witten and Sander \cite{witten1981diffusion}; DLCA is implemented according to the algorithm described in Ref.~\cite{kolb1983scaling}. For each aggregate type (see Fig.~\ref{fig:overview}c), we generate ten distinct realizations with $N=5,~10,~20,~40,~60,~80,~100,~120,\allowbreak~160,~200,~250,~300,~350,~400,~500,~600,~700,~800,~900,\allowbreak~1000$ particles (SAW, DLA, DLCA) or edges (LA). DLA and DLCA are implemented using periodic boundary conditions, which are unraveled before applying KR theory such that only one connected aggregate is considered and the particles in it are only interacting with other particles of itself. 

\paragraph{Material parameters:} We choose a particle radius of $a=5~\mathrm{nm}$ and a center-to-center distance of $15~\mathrm{nm}$. The bulk viscosity is $1~\mathrm{mPa\,s}$ and the membrane viscosity is assumed to be $1~\mathrm{pN \, \mu s/nm}$ \cite{faizi2022vesicle}. 

\paragraph{Random-interior approximation:}
The buffer and convex hull are computed using the \code{buffer} and \code{convex\_hull} implementations of the Python package Shapely \cite{Shapely}. To place particles within a given outline, we use either a Poisson disk or grid-based sampling approach. For Poisson disk sampling, we employ SciPy's \cite{2020SciPy-NMeth} \code{fill\_space} implementation to fill the bounding box of the outline, with the same minimum center-to-center particle distance as used for the aggregation models. From this sample, we then draw $N$ particles that are within the outline. For small aggregates, this approach may fail to yield a sufficient number of particles within the outline even after several sampling attempts. In that case, we superimpose the outline with a square grid with the original lattice spacing and attempt to place $N$ particles in the interior of the outline. With this method, it may become necessary to shift the grid to obtain sufficiently many particles within the outline. We avoid using this method as a default as it may lead to very similar configurations as the original aggregation algorithms.  